\documentclass{spie}

\usepackage{microtype}
\usepackage[separate-uncertainty]{siunitx}
\usepackage{graphicx}
\usepackage[hidelinks]{hyperref}
\usepackage{fullpage}
\usepackage{textgreek}

\usepackage[hang,perpage]{footmisc}

\renewcommand\Affilfont{\normalsize}

\makeatletter
\renewcommand\@author{\ifx\AB@affillist\AB@empty\AB@author\else
      \ifnum\value{affil}>\value{Maxaffil}\def\rlap##1{##1}%
    \AB@authlist\\[\affilsep]\ifx\@behalf\empty\else\Affilfont\@behalf\vspace{\affilsep}\\\fi\AB@affillist
    \else  \AB@authors\fi\fi}
\def\behalf#1{\gdef\@behalf{#1}}
\behalf{}     
\makeatother

\makeatletter
\newcommand\authorcount{17}
\renewcommand\AB@authnote[1]{\ifnum\value{authors}<\authorcount\relax,\fi\textsuperscript{\normalfont#1}}

\makeatother

\begin{document}

\title{The Precursor Small Aperture Telescope (PreSAT)\\CMB polarimeter}

\behalf{With support of the \textsc{Bicep}/\emph{Keck} and CMB-S4 collaborations}

\author[a]{\href{https://orcid.org/0000-0002-4436-4215}{Matthew~A.~Petroff}}
\author[b]{\href{https://orcid.org/0000-0002-9957-448X}{Zeeshan~Ahmed}}
\author[c,d]{James~J.~Bock}
\author[a]{\href{https://orcid.org/0000-0002-3519-8593}{Marion~Dierickx}}
\author[c]{\href{https://orcid.org/0000-0002-3790-7314}{Sofia~Fatigoni}}
\author[b]{\href{https://orcid.org/0000-0001-5268-8423}{David~C.~Goldfinger}}
\author[a]{\href{https://orcid.org/0000-0001-9292-6297}{Paul~K.~Grimes}}
\author[b]{Shawn~W.~Henderson}
\author[b]{\href{https://orcid.org/0000-0002-5215-6993}{Kirit~S.~Karkare}}
\author[a]{\href{https://orcid.org/0009-0003-5432-7180}{John~M.~Kovac}}
\author[d]{Hien~T.~Nguyen}
\author[a]{\href{https://orcid.org/0000-0003-4622-5857}{Scott~N.~Paine}}
\author[a]{Anna~R.~Polish}
\author[e]{Clement~Pryke}
\author[c]{Thibault~Romand}
\author[f]{Benjamin~L.~Schmitt}
\author[g]{Abigail~G.~Vieregg}
\affil[a]{Center for Astrophysics, Harvard \& Smithsonian, Cambridge, MA 02138, USA}
\affil[b]{Kavli Institute for Particle Astrophysics and Cosmology, SLAC National Accelerator Laboratory, Menlo~Park, CA 94025, USA}
\affil[c]{Department of Physics, California Institute of Technology, Pasadena, CA 91125, USA}
\affil[d]{Jet Propulsion Laboratory, Pasadena, CA 91109, USA}
\affil[e]{Minnesota Institute for Astrophysics, School of Physics and Astronomy, University of Minnesota, Minneapolis, MN 55455, USA}
\affil[f]{Department of Physics and Astronomy, University of Pennsylvania, Philadelphia, PA 19104, USA}
\affil[g]{Department of Physics, Enrico Fermi Institute, Kavli Institute for Cosmological Physics, University~of~Chicago, Chicago, IL 60637, USA}

\authorinfo{Further author information: (Send correspondence to M.~A. Petroff)\\E-mail: mpetroff@cfa.harvard.edu}

\maketitle

\vspace{0.4em}

\begin{abstract}
The search for the polarized imprint of primordial gravitational waves in the cosmic microwave background (CMB) as direct evidence of cosmic inflation requires exquisite sensitivity and control over systematics. The next-generation CMB-S4 project intends to improve upon current-generation experiments by deploying a significantly greater number of highly-sensitive detectors, combined with refined instrument components based on designs from field-proven instruments. The Precursor Small Aperture Telescope (PreSAT) is envisioned as an early step to this next generation, which will test prototype CMB-S4 components and technologies within an existing \textsc{Bicep} Array receiver, with the aim of enabling full-stack laboratory testing and early risk retirement, along with direct correlation of laboratory component-level performance measurements with deployed system performance. The instrument will utilize new 95/\SI{155}{\giga\hertz} dichroic dual-linear-polarization prototype detectors developed for CMB-S4, cooled to \SI{100}{\milli\kelvin} via the installation of an adiabatic demagnetization refrigerator, along with a prototype readout chain and prototype optics manufactured with wide-band anti-reflection coatings. The experience gained by integrating, deploying, and calibrating PreSAT will also help inform planning for CMB-S4 small aperture telescope commissioning, calibration, and operations well in advance of the fabrication of CMB-S4 production hardware.
\end{abstract}

\keywords{cosmic microwave background, telescopes, cryogenics}

\section{Introduction}

Anisotropy of the cosmic microwave background (CMB), which consists of light emitted a few hundred thousand years after the Big Bang, encodes a plenitude of information about the early Universe, and measurements of the CMB have resulted in the establishment of \textLambda{}CDM as the standard model of cosmology.\cite{Bennett2013} However, direct evidence of the cosmic inflation paradigm has not yet been found. This paradigm predicts a period of rapid expansion in the very early Universe,\cite{Guth1981} which would result in primordial gravitational waves and a $B$-mode signal in the CMB polarization anisotropy, parameterized by the tensor-to-scalar ratio, $r$.\cite{Kinney1998} The current upper limits on this signal are provided by the \textsc{Bicep}/\emph{Keck} series of CMB experiments,\cite{BK18} the most recent of which are the Stage-III \textsc{Bicep3} and \textsc{Bicep} Array (BA) experiments.\cite{B3, BA} While work on and observations from \textsc{Bicep3} and \textsc{Bicep} Array continue, planning and design work are progressing on the next-generation CMB-S4 experiment, which intends to deploy significantly-greater sensitivity to search for $B$-mode polarization sourced by inflationary primordial gravitational waves.\cite{S4}

\begin{figure}
\centering
\def\svgwidth{0.94\textwidth}
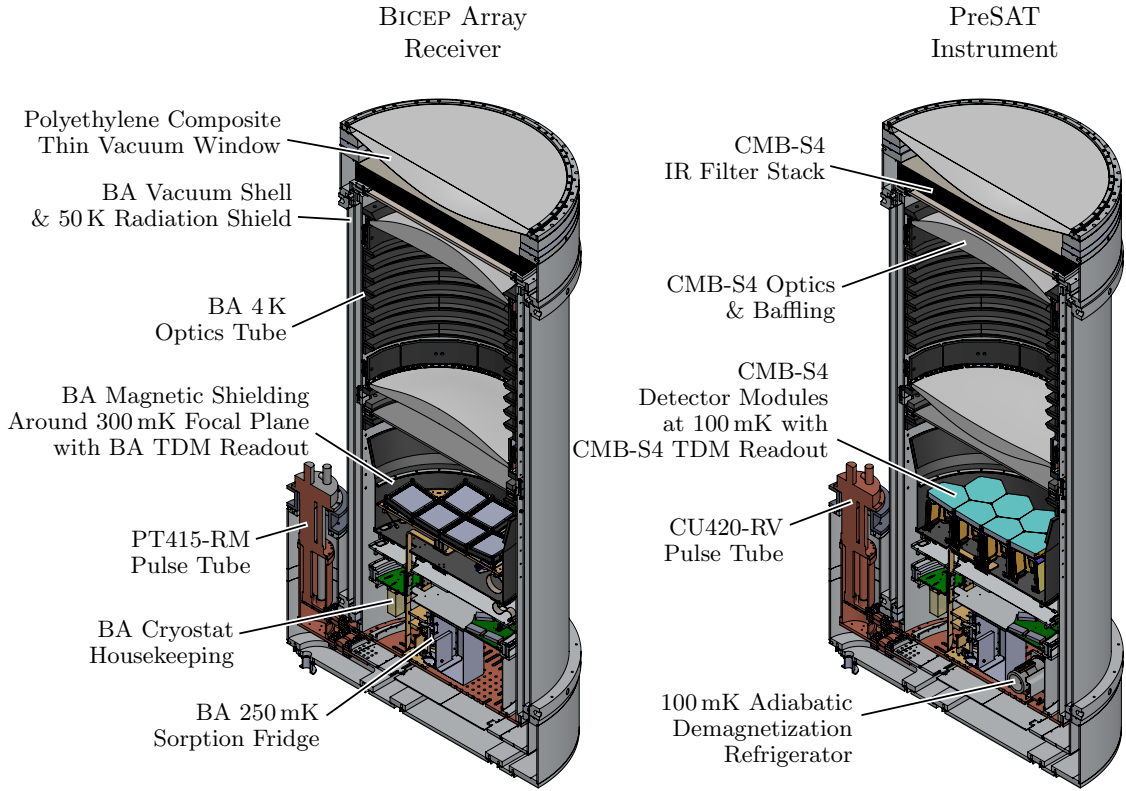
\caption{Experiment overview. Comparison of a \textsc{Bicep} Array (BA) receiver to a mockup of the PreSAT instrument. PreSAT will resemble a BA receiver but will utilize a \SI{100}{\milli\kelvin} adiabatic demagnetization refrigerator (backed by the BA sorption fridge) and feature CMB-S4 prototype optics, detectors, and time-division multiplexing (TDM) readout, among other potential changes.}
\label{fig:overview}
\end{figure}

The Precursor Small Aperture Telescope (PreSAT) experiment is intended to serve as a stepping stone between Stage-III CMB experiments and CMB-S4. The experiment will test prototype CMB-S4 components and technologies within an existing \textsc{Bicep} Array receiver (``BA3''), thereby enabling full-stack laboratory testing and early risk retirement. Furthermore, it will allow for comparative measurements between these new components and existing Stage-III designs and seeks to provide direct correlation of laboratory component-level performance measurements with deployed system performance, with the intention of deploying the completed receiver as the fourth \textsc{Bicep} Array receiver at the South Pole. An overview of the experiment is shown in Figure~\ref{fig:overview}; in comparison to a \textsc{Bicep} Array receiver, detectors, readout, optics, and cryogenics components will be swapped out for either prototype CMB-S4 components or other components necessary to support testing of said prototype components. Among other changes, \textsc{Bicep} Array's square slot-antenna-array-coupled detector modules operating with a \SI{\sim 250}{\milli\kelvin} bath temperature\cite{Schillaci2023} will be at least partially replaced with dichroic hexagonal feedhorn-coupled CMB-S4 detector modules operating with a \SI{100}{\milli\kelvin} bath temperature,\cite{Barron2022} and \textsc{Bicep} Array's heritage single-level time-division multiplexing (TDM) readout\cite{Battistelli2008} will be replaced with CMB-S4's new two-level TDM readout.\cite{Goldfinger2024}

PreSAT aims to help answer outstanding questions that need better answers regarding experiments designed to search for evidence of primordial gravitational waves and the signature of inflation in the CMB, while building on the experience of a currently-working approach. Some of these questions include:
\begin{enumerate}
\item How can we identify and minimize the leading systematics?
\item What are the sources of main-beam mismatch and temperature-to-polarization leakage?
\item What drives 1/$\ell$ noise and limits low-$\ell$ sensitivity?
\item What are the sources of far sidelobes and forebaffle coupling?
\item What drives data cuts and hurts observing efficiency?
\item How can we create an efficient cryogenic architecture?
\item How can we verify optics focus in the lab?
\item What are the specific goals of field testing that will predict key performance?
\end{enumerate}
Answering these questions and addressing the issues they potentially uncover will allow for the construction of more-sensitive and efficient instruments and thus advance the motivating science goals.

The remainder of this proceedings paper is structured as follows. First, optics are discussed in Section~\ref{sec:optics}. Then cryogenics are discussed in Section~\ref{sec:cryo}, followed by discussion of magnetic shielding and pickup in Section~\ref{sec:mag}. Finally, field deployment is discussed in Section~\ref{sec:deployment}, and conclusions are presented in Section~\ref{sec:conclusions}.

\section{Optics}
\label{sec:optics}

\begin{figure}
\centering
\includegraphics[width=0.8\textwidth]{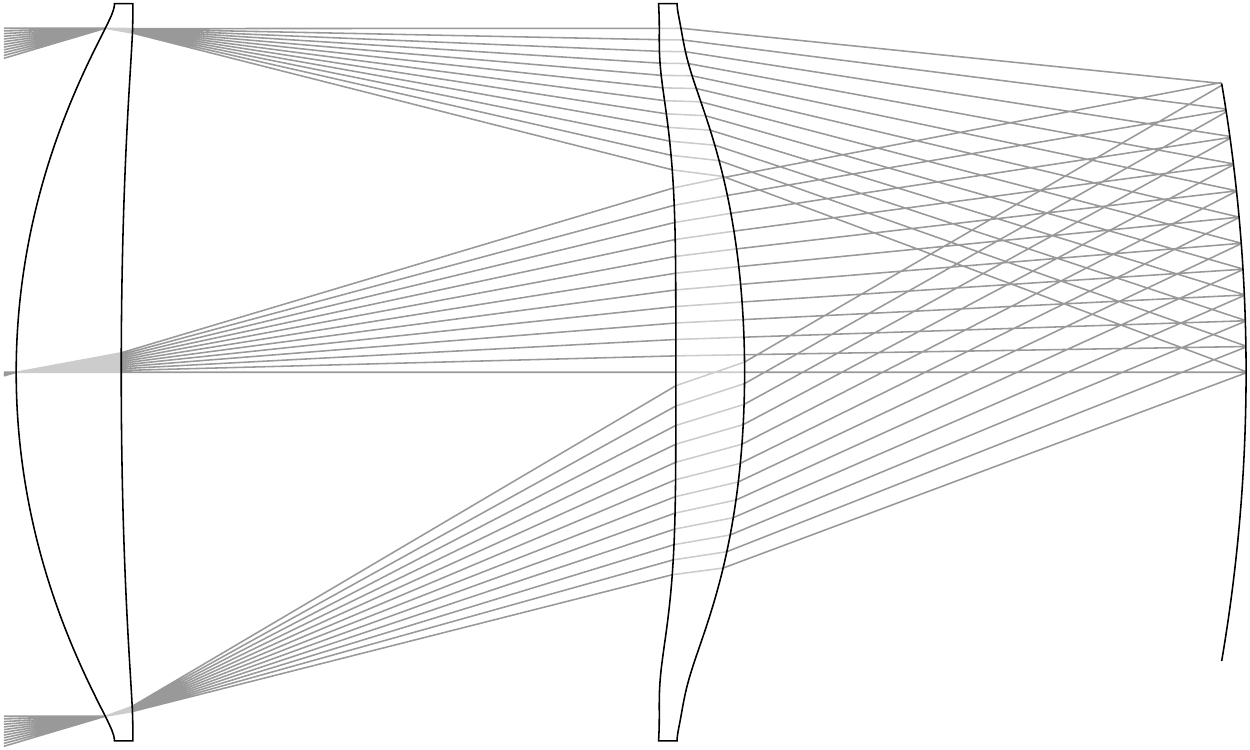}
\caption{Ray trace of a potential optics design incorporating, from left to right, two high-density polyethylene (HDPE) lenses and a curved focal surface with a radius of \SI{1400}{\milli\meter}. Flat filter elements are not included in this diagram.}
\label{fig:optics}
\end{figure}

The PreSAT instrument will be outfitted with a prototype CMB-S4 optics stack. This stack will resemble that of existing \textsc{Bicep} Array receivers and consists of a pair of high-density polyethylene lenses, a series of filters to reduce thermal loading on the cryogenic stages, and a thin vacuum window.\cite{Eiben2022} Although based on similar design principles, the lens prescriptions will be an entirely new design, and all elements will use new anti-reflection coating recipes to account for the wider bandwidth of the receiver's dichroic detectors; a ray trace of a current candidate optics design is shown in Figure~\ref{fig:optics}. While the lenses are intended to match what will be used for CMB-S4 as closely as possible, the thermal filtering will differ as the lenses will be cooled to \SI{4}{\kelvin} instead of \SI{1}{\kelvin}, as is currently planned for CMB-S4; infrared filtering and its effect on cryogenics will be discussed in Section~\ref{sec:cryo}.

For in-lab testing, this optics stack will be used with both prototype feedhorn-coupled CMB-S4 detector modules\cite{Barron2022} and heritage slot-antenna-array-coupled \textsc{Bicep3} detector modules,\cite{Hui2016} enabling direct comparative testing. Besides standard in-lab verification testing such as near-field beam mapping and Fourier-transform spectrometer measurements, this combination of prototype optics and different generations of detector modules will allow us to probe for answers to outstanding questions regarding the sources of beam-related systematics, in particular the sources of far sidelobes and main-beam mismatch.

In order to reduce systematic errors caused by far sidelobes, CMB telescopes often use absorptive forebaffles to terminate these sidelobes. Although this is an effective mitigation strategy, this warm termination increases detector loading, thereby reducing sensitivity; this tradeoff motivates finding ways to reduce these sidelobes and terminate as much stray light as possible within the cold receiver. Diffraction calculations of a beam through the telescope's aperture only explain a small fraction of this forebaffle coupling on experiments such as \textsc{Bicep3} and \textsc{Bicep} Array, so the source of the majority of the coupling is unclear. Fortunately, forebaffle coupling can be straightforwardly measured in the lab using detector-loading measurements in combination with an absorptive forebaffle with a liquid-nitrogen cold load placed on top, so elements of the cryostat and optics stack, e.g., the aperture stop and absorptive baffling on the walls of the cold optics tube, can be replaced and iterated on in an attempt to identify and reduce the sources of this excess coupling. This forebaffle coupling will also be compared between \textsc{Bicep3} and prototype CMB-S4 detector modules, which utilize significantly-different methods to couple to incoming radiation; additionally, different feedhorn designs---with different detector beam sizes and thus different edge tapers at the aperture---will be used with the prototype CMB-S4 detector modules as an additional source of comparison. While no similarly-straightforward method of probing main-beam mismatch exists, efforts will also be made to better understand its sources and also reduce it.

Beyond existing pre-deployment in-lab verification testing such as bandpass and optical-efficiency measurements and near-field beam mapping, there is also a desire to characterize the far-field beam, both to further verify the optical design and fabrication and to verify the focus of the optical system, such that it can be fine-tuned prior to field deployment, if necessary. Given the impracticality of in-lab far-field beam mapping with a source hundreds of meters away, both phase and intensity information on the beam must be recorded in the near field with holography, such that the far-field beam can be estimated. Although this has previously been done for CMB telescope receivers using a cryogenic coherent receiver temporarily installed at the focal surface of the receiver,\cite{Chesmore2022} this requires both a dedicated cryostat cold run and a specialized detector, readout chain, and wiring for this specific test, and any inadvertent changes to the receiver or misplacement of this detector would invalidate the results. Thus, this holography would ideally be done with bolometers installed in the receiver in their field-ready configuration, but this requires a different approach, as bolometers do not directly measure phase. Instead, two phase-locked sources can be used outside the receiver. One source is kept fixed, while the other is scanned across the aperture, such that the bolometers measure the intensity of the interference fringes; both sources must be coupled to the detectors to be measured. Such an approach has previously been used in the submillimeter for kinetic inductance detectors using an aperture-filling beam splitter\cite{Yates2020} and for CMB bolometers in a catadioptric optical system, with the fixed source bypassing the reflective optics to fill the aperture.\cite{Nakano2023}

\begin{figure}
\centering
\includegraphics[width=4in]{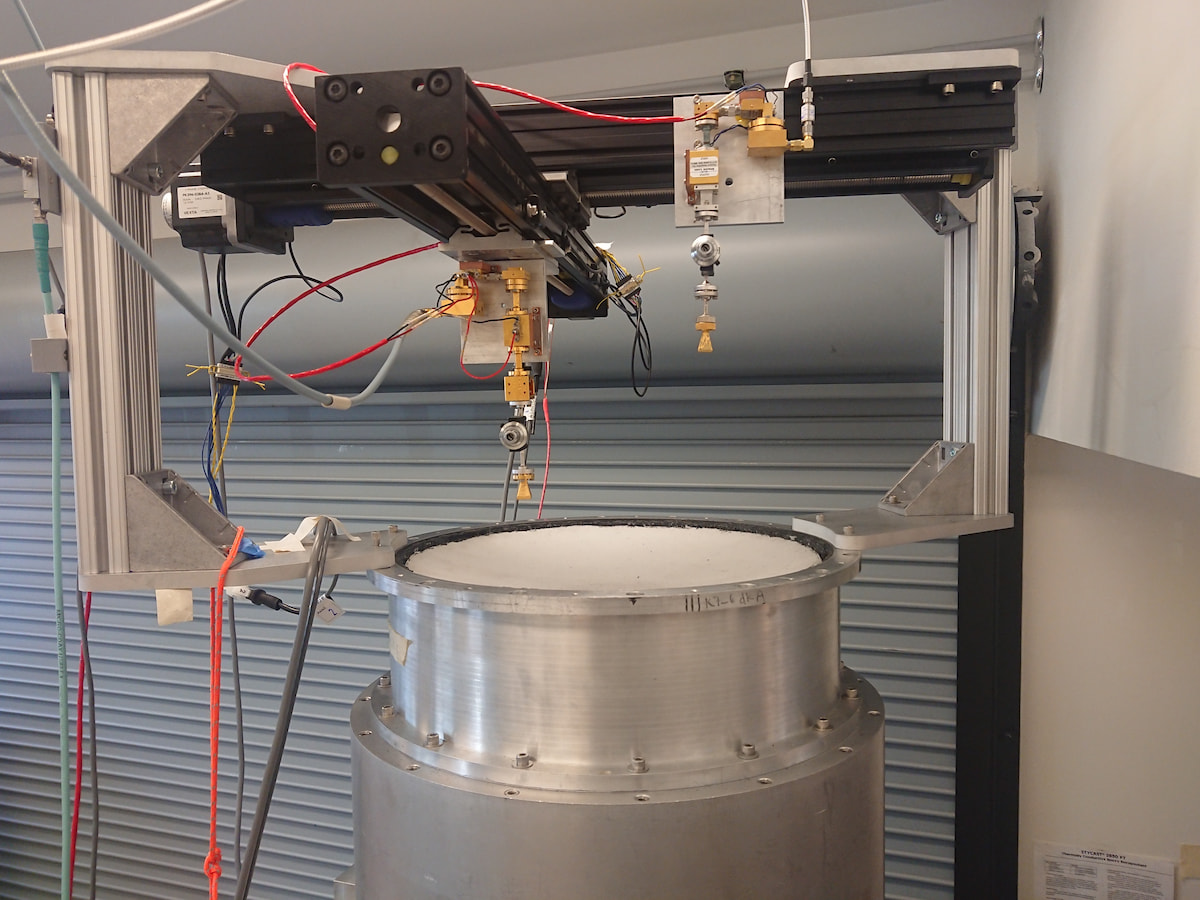}
\caption{Holography setup installed on a \emph{Keck} Array receiver. To the left is the movable source on an x-y translation stage, while the fixed source is on the right.}
\label{fig:holography}
\end{figure}

However, PreSAT has no mirrors to bypass, and a \SI{>70}{\centi\meter} beamsplitter necessary to fully cover the instrument's full vacuum window is difficult to construct. Fortunately, neither is necessary; with a wide beam, the fixed source can be placed off to the edge of the vacuum window and still couple to the full focal plane. Such a holography system has been constructed at \SI{220}{\giga\hertz} and is currently being tested using a \emph{Keck} Array receiver, as shown in Figure~\ref{fig:holography}; there are plans to use it to verify the focus of \textsc{Bicep} Array's 220/\SI{270}{\giga\hertz} BA4 receiver, once the holography system is validated. While characterizing the far-field beam to the precision necessary for cosmological analysis with such an approach is challenging, measuring the focus of the optical system needs significantly less precision and thus provides an actionable near-term goal. This system will later be extended to operate at additional frequencies to be used for PreSAT and later CMB-S4.

\section{Cryogenics}
\label{sec:cryo}

The overall power draw of a contemporary CMB experiment is driven by cryogenics, specifically the power draw of pulse-tube cryocoolers, so the implementation of an efficient cryogenic architecture is necessary to minimize the experiment's energy usage. At the South Pole, there is a hard limit to the maximum power draw of experiments due to the sizing of the South Pole Station's power plant, as well as the logistical costs of transporting fuel to the Pole to run it.\cite{spsmasterplan} For telescopes deployed in Chile, a hard power cap does not exist, but it is still prudent to minimize energy usage, to reduce the logistical and carbon footprints of the experiment. While these footprints can also be reduced through the use of renewable energy generation capacity,\cite{Babinec2024} reducing energy consumption is still critical.

For a South Pole deployment, CMB-S4 would likely need to fit within the power budget of the existing Stage-III CMB experiments. The current South Pole plan for CMB-S4 calls for nine small aperture telescope (SAT) optics tubes,\cite{aoa} while the current \textsc{Bicep3} and \textsc{Bicep} Array experiments have a total of only five similarly-sized optics tubes. Thus, improved efficiency is critical for such a deployment to be feasible. Although this is the preferred configuration for CMB-S4, the current inability of the National Science Foundation to commit support for a South Pole component of CMB-S4\cite{bpa} may make a Chile-only alternative necessary.

For a Chile-only deployment, the CMB-S4 analysis of alternatives calls for at least 27 small aperture telescope (SAT) optics tubes and at least three large aperture telescopes (LATs).\cite{aoa} For a naive implementation, one could extrapolate from telescope designs for the Simons Observatory, which is currently under construction at the proposed CMB-S4 site in Chile. The Simons Observatory SATs each contain one optics tube cooled by a pair of Cryomech\footnote{Bluefors Cryocooler Technologies Inc., Syracuse, NY 13211, USA; \url{https://bluefors.com/}} PT420-RM pulse tubes,\cite{Galitzki2024} and the Simons Observatory LATs are cooled by a pair of PT420-RM pulse tubes and a pair of Cryomech PT90-RM pulse tubes.\cite{Bhandarkar2022} Per the manufacturer specifications, each PT420 draws \SI{12.5}{\kilo\watt}, while each PT90 draws \SI{5.0}{\kilo\watt}. Although this power draw is not unreasonable for an experiment the size of Simons Observatory, extrapolation to a proposed Chile-only CMB-S4 configuration results in a load of \SI{780}{\kilo\watt} for just pulse-tube cryocoolers, which would push the overall experiment's power usage toward \SI{1}{\mega\watt}. While installing the necessary generation capacity at the off-grid site is technically feasible, significantly-reduced power usage would be preferred.

With a given thermal load, increasing cryocooler efficiency will reduce power usage. Pulse-tube cryocooler manufacturers generally specify cooling performance at a specific cold-head temperature, with the pulse tube in a vertical orientation. However, CMB telescopes, particularly those with a larger range of boresight rotation, can operate with the pulse tube in an orientation that is tilted significantly off vertical. Additionally, the desired cold-head temperature is not necessarily the same as the manufacturer's specification. Helium charge pressure and rotary-valve frequency can both be adjusted, as can compressor frequency in the case of inverter-driven variable-frequency compressor models. The impedance and volume of the helium lines between the pulse tube and its helium compressor---determined by line diameter and length, flow restrictions such as rotary unions, ballast volumes, etc.---affect cooling efficiency. Finally, there are multiple manufacturers of pulse tubes and multiple models commercially available, each with different performance. Thus, comprehensive testing is required, in a range of orientations, to better understand pulse-tube performance and efficiency. Such testing is planned for PreSAT. Cryomech PT410-RM and PT415-RM pulse tubes will be evaluated, as will a Boston Cryogenics\footnote{Boston Cryogenics LLC, North Billerica, MA 01862, USA; \url{https://www.boscryo.com/}} CU420-RV pulse tube. The cryocoolers will be outfitted with thermometers, heaters, and helium pressure sensors and evaluated in a range of operating conditions and orientations, all while energy consumption is monitored, with testing performed in a purpose-built test cryostat.

The other method of reducing cryocooler energy usage is to reduce heat load on the cryocooler. While the exact amounts of heat load are specific to individual instrument designs, some design principles for reducing loads, e.g., through design of radiation shields, multi-layer insulation (MLI), and mechanical supports, are generally applicable to all cryostats. However, telescopes have the additional requirement of a large aperture for observing the sky, which cannot be covered with traditional radiation shields or MLI. Thermal loading on all stages of the cryostat through this aperture must be controlled via filters that are low loss and transparent at millimeter wavelengths. These filters can operate through reflection, absorption, or scattering of thermal radiation. Reflective filters are typically polypropylene patterned with a conductive capacitative grid, while absorptive filters are typically constructed from alumina, polytetrafluoroethylene (PTFE), or nylon. While previous-generation CMB telescopes typically used reflective filters, these have largely been replaced with radio-transparent multi-layer insulation (RT-MLI),\cite{Choi2013} which consists of a stack of low-refractive-index dielectric layers that absorb and re-emit thermal radiation, similar to traditional MLI. However, different experiments have used different materials for this filter stack, with some using polyethylene foam and others using polystyrene foam, with other candidates being nylon foam or expanded PTFE. Even for conceptually-simple absorptive filters, there are size-dependent effects due to thermal conductivity, which has driven a switch from PTFE to alumina as telescope apertures have increased in diameter. Thus, additional testing, including testing of full-size filters, is needed to better optimize these filter choices, and such testing is planned for PreSAT. Filtering at the warmest stages drives heat load on the pulse tube cryocooler and is thus important for energy usage, while the lenses also play a role in the radiative load on a cryostat's coldest stages. By utilizing a prototype CMB-S4 SAT optics stack, PreSAT aims to retire risk by evaluating radiative loading through the full range of cryostat temperature stages.

For sub-kelvin cooling, \textsc{Bicep} Array receivers use a three-stage \textsuperscript{3}He/\textsuperscript{4}He sorption fridge provided by CEA/SBT\footnote{French Alternative Energies and Atomic Energy Commission, Low Temperature Systems Department;\\\url{https://www.d-sbt.fr/en}} to reach temperatures of \SI{250}{\milli\kelvin}, but a \SI{100}{\milli\kelvin} temperature stage is required for CMB-S4 detectors. To accommodate this colder temperature stage, the \textsc{Bicep} Array receiver used for PreSAT will be modified. An adiabatic demagnetization refrigerator (ADR) will be installed, backed by the \SI{\sim 340}{\milli\kelvin} second stage (``intercooler'') of the sorption fridge, which will replace (or extend) the \SI{250}{\milli\kelvin} stage of the sorption fridge. This single-shot ADR, also produced by CEA/SBT, is based on the SPICA/BLISS cryogenic demonstrator.\cite{Prouve2015} An initial ADR unit has been evaluated in a testbed at Caltech and found to have \SI{\sim 48}{\hour} of hold time with \SI{\sim 2}{\micro\watt} of total heat load;\cite{Romand2024} this heat load is similar to what is expected for PreSAT, based on heat-load measurements of the \SI{250}{\milli\kelvin} stage of \textsc{Bicep} Array's \SI{150}{\giga\hertz} BA2 receiver in the field. The final ADR unit for PreSAT is currently being manufactured and is expected to be delivered in the coming months. It will be integrated into the PreSAT receiver once it has been delivered.

\section{Magnetic shielding and pickup}
\label{sec:mag}

CMB-S4 plans to use transition-edge-sensor (TES) bolometers read out using time-division multiplexing (TDM).\cite{Barron2022} This readout scheme utilizes superconducting quantum interference devices (SQUIDs), which are extremely sensitive to magnetic fields. Thus, extensive magnetic shielding is necessary, both at the receiver and detector-module levels. This shielding typically consists of several layers of both high-magnetic-permeability (high-\textmu) and superconducting materials. For receiver-level shielding, \textsc{Bicep} Array uses a cylindrical Amuneal A4K high-\textmu{} shield\footnote{Amuneal Manufacturing Corp., Philadelphia, PA 19124, USA; \url{https://www.amuneal.com/}} on its \SI{50}{\kelvin} radiation shield and a superconducting niobium flared cup shield cooled to near \SI{350}{\milli\kelvin}, although other superconducting shields such as tin--lead plating on copper\cite{Galitzki2024} are possible. In order to set a magnetic-shielding performance baseline for CMB-S4, this shielding setup needs to be better characterized and understood.

\begin{figure}
\centering
\includegraphics[height=3.6in]{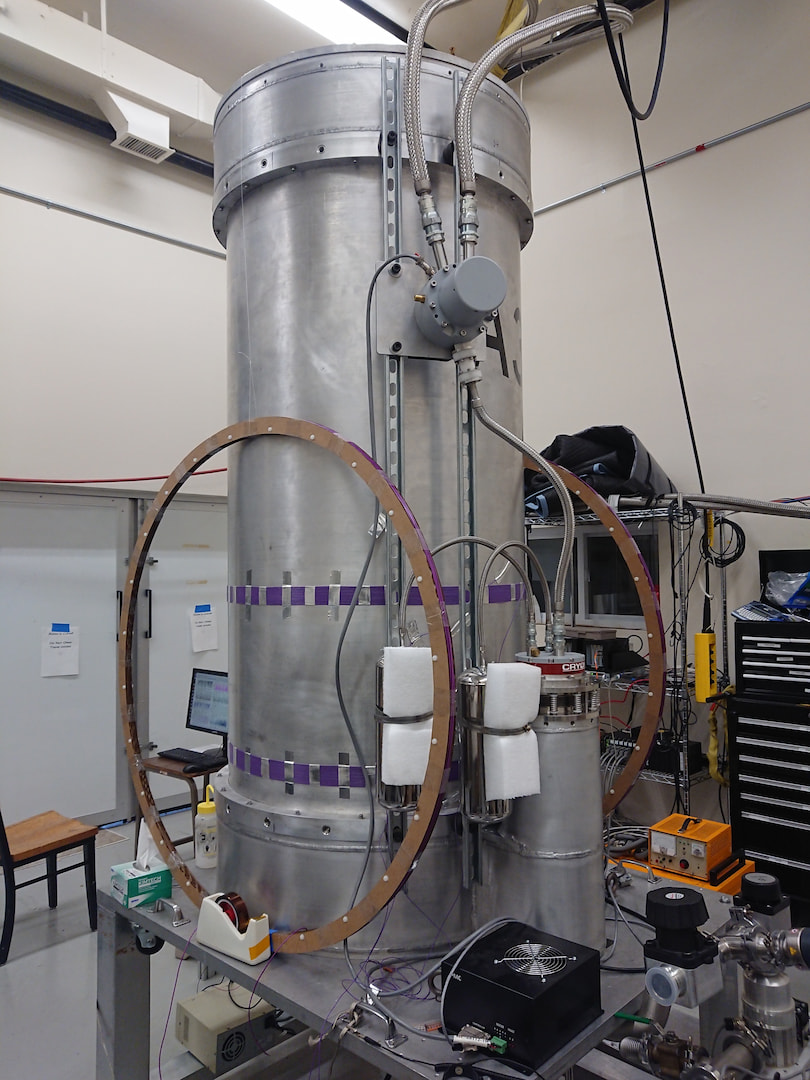}\hfill
\includegraphics[height=3.6in]{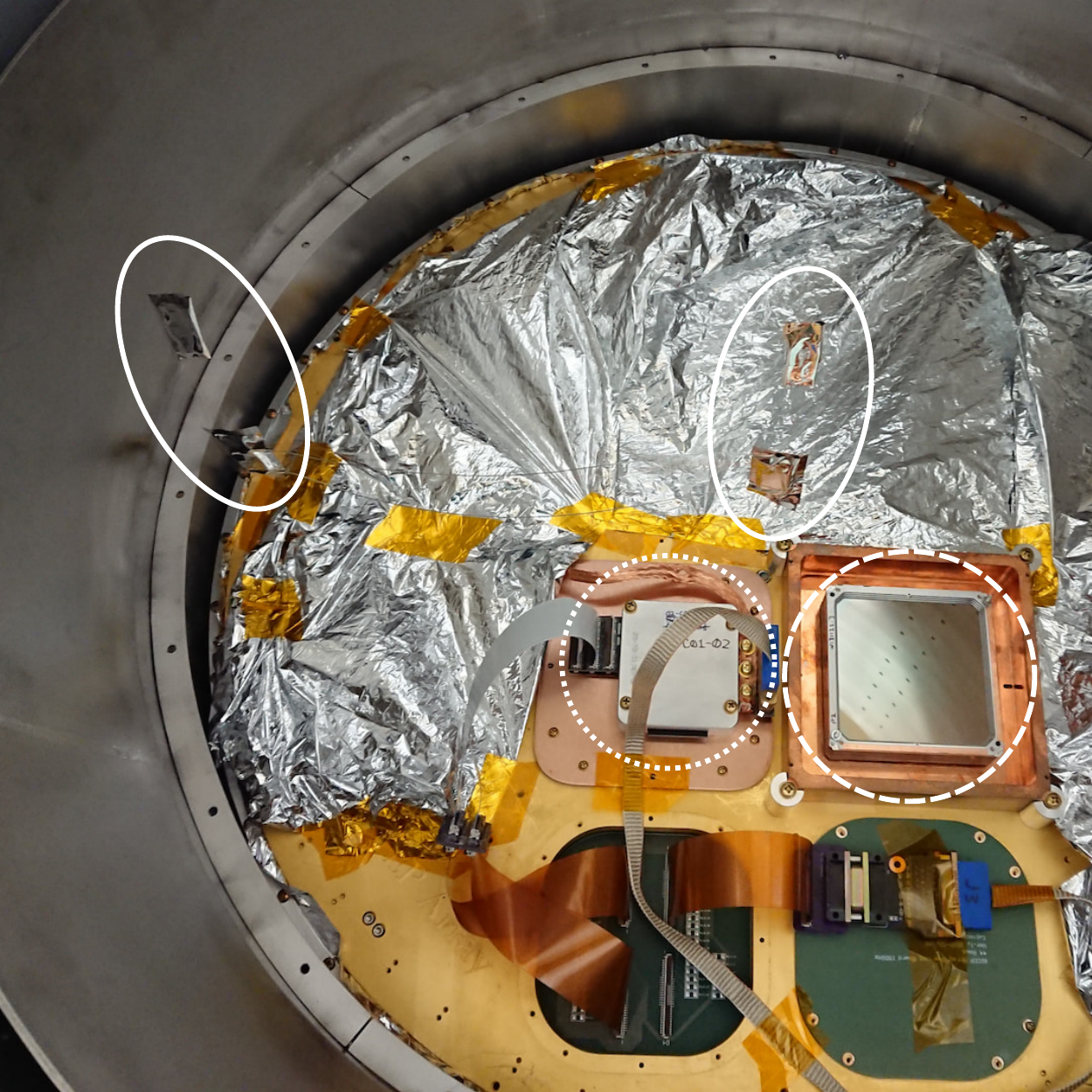}
\caption{Magnetic testing setup. Left: Helmholtz coils installed on PreSAT receiver. Right: Magnetometers (in solid circles) installed inside cryostat. A \textsc{Bicep3} detector module (in dashed circle) and a CMB-S4 SQ1 test module (in dotted circle) are installed in the focal plane. The niobium cup is visible around the left and top edges.}
\label{fig:magnetometers}
\end{figure}

The BA3 receiver was outfitted to test the existing \textsc{Bicep} Array magnetic shielding. Both \SI{0.8}{\meter} diameter axial and \SI{1.2}{\meter} diameter transverse Helmholtz coils\footnote{The coils used to produce transverse fields are spaced too far apart to be true Helmholtz coils, due to space constraints limiting their diameter to $\sim$1.35 times their spacing.} were installed; these are driven using a constant-current linear drive amplifier connected to a function generator and can produce magnetic fields up to around \SI{500}{\micro\tesla}, several times the strength of the geomagnetic field, at excitation frequencies down to DC. To measure magnetic fields, a set of four three-axis magnetometers was installed inside the niobium cup; these magnetometers were placed along the cup's axis of revolution and at the edge, in both the approximate axial position of the \textsc{Bicep} Array detector modules and approximately \SI{10}{\centi\meter} closer to the cup's opening. Each magnetometer module consists of a \textsc{Memsic}\footnote{\textsc{Memsic} Semiconductor Co., Ltd., Tianjin, China; \url{https://www.memsic.com/}} MMC5983MA three-axis anisotropic magneto-resistive magnetometer, readout circuitry, and an analog temperature sensor and operational amplifier to servo the module to approximately \SI{250}{\kelvin}. These modules are covered in low-emissivity tape and suspended by a total of four approximately meter-long \SI{100}{\micro\meter} diameter enameled wires from the top of the cryostat's \SI{4}{\kelvin} tube to minimize thermal loading. Digital readout is performed via an I\textsuperscript{2}C bus shared among the four modules. Between the four modules and resistive losses in the wiring, total power usage is approximately \SI{200}{\milli\watt}, with the majority of the power going to the temperature-control circuit. This additional heat load is well within the margin of the \SI{4}{\kelvin} stage of the pulse tube, but the sub-kelvin sorption fridge could not be cycled with the magnetometer modules powered on, due to the excess radiatively-coupled load. Photographs of the Helmholtz coils and magnetometers are shown in Figure~\ref{fig:magnetometers}.

Data were recorded with the Helmholtz coils driven by both \SI{0.1}{\hertz} and \SI{0.01}{\hertz} sine waves, with the latter used as an approximation for the DC response. Both the axial and transverse Helmholtz coils were driven independently, each at four different current levels, and measurements were made with both the niobium cup superconducting and with it normal. These data, of magnetic field strength and direction in four locations, will allow for validation of receiver-level magnetic-shielding simulations. Additionally, comparing the data with the niobium cup superconducting and with it normal will allow the shielding level to be factorized into the separate contributions of the high-\textmu{} cylinder and the superconducting niobium cup.

Once these data were collected, the magnetometers were powered off, which allowed the sub-kelvin sorption fridge to be cycled and the focal plane to be cooled to below \SI{300}{\milli\kelvin}. This allowed the \textsc{Bicep3} detector module and CMB-S4 SQ1 test module installed on the focal plane to be operated. By using the Helmholtz coils in identical configurations to what was previously measured with the magnetometers, repeated measurements with known magnetic field strengths were made. This allowed the magnetic sensitivity of the two modules to be evaluated in known magnetic fields. The CMB-S4 SQ1 test module utilizes an aluminum case, so measurements were made above and below the case's superconducting transition; by combining the magnetic field strength measurements with the pickup seen in the SQ1 data taken with the aluminum case normal, the effective magnetic cross section of the NIST MUX15b SQ1 chips can be calculated.

\section{Field Deployment}
\label{sec:deployment}

While much can be learned from in-lab testing, laboratory conditions are fundamentally different from a field-deployed configuration. Due to the much-warmer radiative environment, the TES bolometers operate on a different superconducting transition in the lab than they do on the sky, and the laboratory presents a much-noisier RF environment. Environmental conditions are different, with the laboratory environment unable to replicate the extreme cold the vacuum window is exposed to in the field. A stationary cryostat does not experience the vibrations and tilts associated with a scanning telescope mount, and even if a receiver were to be mounted on a duplicate mount in the lab, this would not provide the same magnetic field conditions. Finally, the far field of the telescope's optical system cannot be directly probed in the lab.

Thus, in addition to providing in-lab integrated receiver-level testing, PreSAT is ultimately intended to be deployed to the South Pole, as the fourth \textsc{Bicep} Array receiver. This will provide invaluable integrated testing of the completed receiver in its deployed configuration and provide direct comparative testing against Stage-III experiments. This will allow for subtle systematic errors to be probed for and fully characterize how the telescope's different systems interact with each other and with the on-sky observing environment. In particular, beam response and near and far sidelobes and their interaction with the terrain (and atmosphere) through the shielding provided by the telescope's ground shield and forebaffle are complex. The effective temperature of the terrain surrounding the telescope can vary on order \SI{100}{\kelvin} on the degree scales of interest. The CMB-S4 science goal of placing an upper limit of $r < 0.001$ at 95\% confidence or detecting $r = 0.003$ at high confidence\cite{S4forecast} requires \SI{<10}{\nano\kelvin} uncertainties at degree scales and thus requires systematics from ground pickup to be suppressed by better than ten orders of magnitude in amplitude (twenty orders in power). No matter how much modeling and simulation is done, how a telescope will perform on the sky can only be fully determined by actual observations in its deployed configuration. Furthermore, the fraction of data cut and amount of filtering necessary to produce maps and spectra with sufficiently low systematic errors can only be determined by actual observations, and these factors can have a significant impact on an experiment's final sensitivity. Thus, PreSAT aims to provide an avenue for this critical risk-retirement step for CMB-S4.

\section{Conclusions}
\label{sec:conclusions}

PreSAT seeks to answer outstanding questions with regard to technical developments and to both sources of and minimization of systematic errors to inform future development of CMB-S4 SATs. In order to answer these questions, PreSAT is intended to provide a platform for full-stack receiver-level testing, including direct comparative testing, and risk retirement by integrating prototype CMB-S4 technologies into a \textsc{Bicep} Array receiver. Optics, cryogenics, and magnetic shielding are currently undergoing active testing, and direct comparative testing will continue to evolve as additional prototype CMB-S4 hardware is received and integrated. After laboratory testing is concluded, PreSAT is intended to be deployed to the South Pole as the fourth \textsc{Bicep} Array receiver, providing definitive on-sky testing and comparisons with Stage-III instruments.

\acknowledgments

We acknowledge the National Science Foundation Division of Astronomical Sciences for their support of PreSAT under Grant Number 2216223, as well as contributions of hardware, time, and expertise from the \textsc{Bicep}/\emph{Keck} and CMB-S4 collaborations.

\bibliography{paper.bib}
\bibliographystyle{spiebib}

\end{document}